# Tree Models Machine Learning to Identify Liquid Metal based Alloy Superconductor

Chen Hua[1,2], Jing Liu[1,2,3]*

[1]*Key Lab of Cryogenic Science and Technology, Technical Institute of Physics and Chemistry, Chinese Academy of Sciences, Beijing 100190, China.*

[2]*School of Future Technology, University of Chinese Academy of Sciences, Beijing 100049, China.*

[3]*School of Biomedical Engineering, Tsinghua University, Beijing 100084, China.*

*Correspondence: jliu@mail.ipc.ac.cn

## Abstract

Superconductors, which are crucial for modern advanced technologies due to their zero-resistance properties, are limited by low $T_c$ and the difficulty of accurate prediction. This article made the initial endeavor to apply machine learning to predict the critical temperature ($T_c$) of liquid metal (LM) alloy superconductors. Leveraging the SuperCon dataset, which includes extensive superconductor property data, we developed a machine learning model to predict $T_c$. After addressing data issues through preprocessing, we compared multiple models and found that the Extra Trees model outperformed others with an $R^2$ of 0.9519 and an RMSE of 6.2624 K. This model is subsequently used to predict $T_c$ for LM alloys, revealing $In_{0.5}Sn_{0.5}$ as having the highest $T_c$ at 7.01 K. Furthermore, we extended the prediction to 2,145 alloys binary and 45,670 ternary alloys across 66 metal elements and promising results were achieved. This work demonstrates the advantages of tree-based models in predicting $T_c$ and would help accelerate the discovery of high-performance LM alloy superconductors in the coming time.

## 1. Introduction

Superconductors exhibit many fascinating properties, including the transmission of high currents without the generation of heat due to zero resistance. These properties render them a promising candidate for advanced devices, including power transmission systems, quantum computers, and micromagnetic resonance devices. Nevertheless, the extensive utilization of superconductors has met two considerable obstacles. (1) The conduction currents of a superconductor are zero-resistance exclusively below the $T_c$. It is typical for superconductors to be cooled to temperatures approaching or below the boiling point of nitrogen (77 K) in order to exhibit zero-resistance characteristics. (2) The theory for predicting $T_c$ remains an open question that has puzzled the researchers since the discovery of superconductivity by Onnes in Leiden in 1911. This illustrates the significance of $T_c$ as a crucial parameter in superconductor applications, as it defines the operational limits. The $T_c$ of different superconductors varies according to their chemical composition.

Interestingly, the first ever superconductor was discovered in mercury (Hg), a type of liquid metal (LM, metals with a melting point at or near room temperature), but its environmental toxicity has restricted the practical utility. Subsequently, study on superconductors has predominantly shifted to other superconducting materials, such as copper base superconductors, iron base superconductors, and hydrides etc. For many decades, there is a paucity of study pertaining to the superconducting properties of LM alloys. With the recent emergence of room temperature LM science and technology, many unknown properties of LM are being increasingly discovered, among which the superconductivity shown by Ga-based alloys has especially attracted attention from researchers[1]. Ga-based alloys represent a type of low-melting-point metals that exhibit a number of advantageous characteristics, including room-temperature fluidity, non-toxicity, ease of manufacturing, rapid patterning[2] and self-healing ability[3]. The fabrication of such superconducting films and circuits is a relatively straightforward and rapid process, rendering it an appropriate material in making superconducting flexible devices. As it was noticed, the conventional preparation and processing techniques employed in the production of superconductors are generally inefficient. Despite recent advances in the manufacture of superconducting electronic devices utilizing lithography technology, the costs remain relatively high. Given the advantages offered by manufacturing technology, exploring LM alloy superconductors with higher $T_c$ represents a significant opportunity for advancing the superconductors science and technology.

Although superconductors have been studied for more than a century, there is so far no complete theory that can fully explain the mechanism, and the discovery of new superconductors still relies on intuition and tremendous trials and errors based on experience. However, superconductors take a lot of time and money to prepare, process and test, and the high cost of such trial-and-error method has led researchers to change their thinking. Once a candidate superconductor is identified, its $T_c$ can be roughly estimated using the McMillan and Allen-Dynes equation based on BCS theory, but the calculation process usually relies on experimental data or empirical estimates to determine the parameters in the equation, which increases the uncertainty of the calculation results. With the development of quantum mechanics and computer technology, density functional theory for superconductors[4,5](SCDFT) or Eliashberg theory[6] can more fully account for the complexity of electron-phonon coupling and Coulomb interaction, and show greater accuracy in predicting $T_c$. However, for unconventional superconductors, such as copper oxide superconductors, iron-based superconductors and heavy fermion superconductors, the above methods may not be applicable because the mechanism of these materials might not be caused by electron-phonon coupling. At the same time, in the prediction of $T_c$, if we start from the perspective of theoretical calculation and search for suitable superconductors one by one through the combination of many materials, the process will be very long, considering the complexity of the calculation system and the limited computing power.

Therefore, there is an urgent need to develop $T_c$ prediction schemes with comparable accuracy to theoretical methods while being computationally efficient. As an alternative strategy to tackle the problem of $T_c$ theoretical calculation, machine learning has displayed the advantages of low computational cost, short execution time and accurate prediction, and can better balance accuracy and efficiency. As machine learning are data-driven, and a large amount of $T_c$ data has been accumulated over the years, researchers have started to use machine learning to discover new superconductors and predict their $T_c$. The present study revolves around the SuperCon dataset (https://doi.org/10.48505/nims.3739), which is currently the largest and most comprehensive superconductor datasets, it contains the known information of superconductors from the experiment and journal publications.

In this study, after pre-processing the SuperCon dataset, the formula is one-hot coded and the tree-based model is used for data regression to train a model for predicting the $T_c$ of LM alloy superconductors ($R^2$=0.9519, RMSE=6.2624), which is in good agreement with the existing experimental data. In LM alloys that can be used for printing devices, we use trained models to find candidates with high $T_c$ values to provide candidate materials for promoting the development of printed superconductors, revealing $In_{0.5}Sn_{0.5}$ as having the highest $T_c$ at 7.01 K. It is important to note that our model does not determine whether the material is a superconductor, it only gives predictions of the $T_c$. Furthermore, we successfully extended the prediction to 2,145 alloys binary and 45,670 ternary alloys across 66 metal elements. This work illustrates the advantages of tree-based models in predicting $T_c$ and accelerates the development of high-performance LM alloy superconductors.

## 2. Methods

The prediction of the $T_c$ of LM alloy superconductors by machine learning necessitates the consideration of two key aspects:

(1) The collection and preprocessing of dataset.

(2) The adoption of suitable algorithms for dataset to develop model.

The process can be visually understood using the flowchart shown in **Figure 1**. From the dataset as the starting point of the process, the data is pre-processed, including deletion, correction, replenishment and description. The processed data is classified in two modes. One is randomly divided into a training set and a validation set according to a certain ratio for model training and model evaluation. The other is randomly and equally divided into five parts of the dataset, each part as a validation set and the remaining four parts as training, and the training is repeated five times for cross-validation to show the effect of the dataset division on the model and to evaluate the stability of the model performance. The most stable model obtained through cross-validation is taken as the optimal model and evaluated from three perspectives of RMSE, $R^2$ and accuracy using the divided training and validation sets to demonstrate the model performance. The model can then be used as a prediction function from formula input to $T_c$ output.

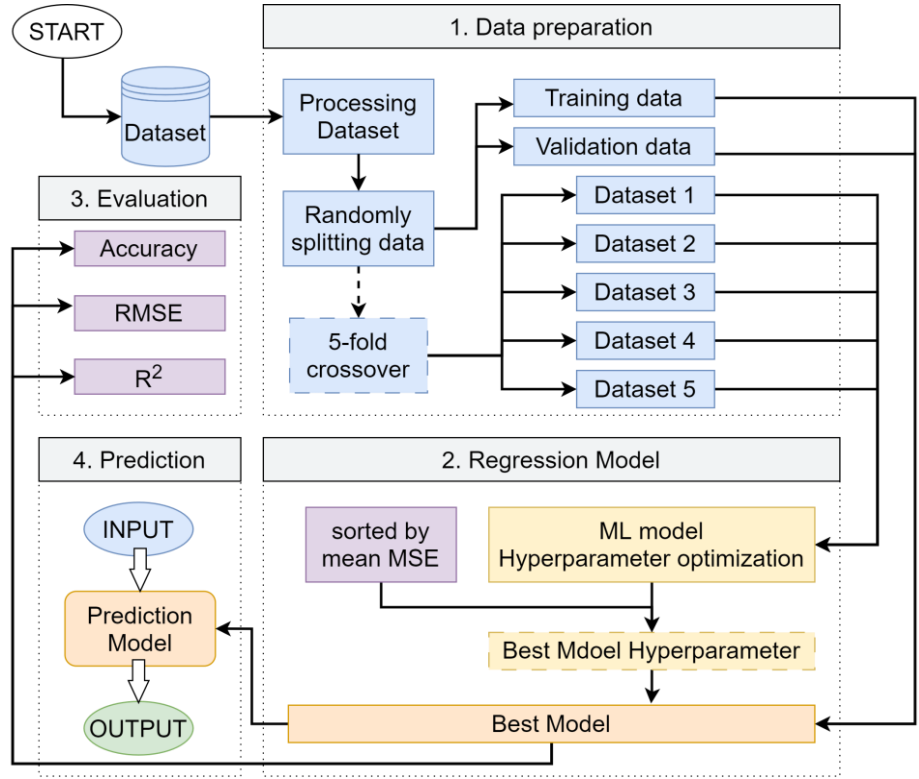

**Figure 1. Workflow diagram of machine learning models**

## 1) Data preprocessing

In this study, the SuperCon dataset **20240322_MDR_OAndM.txt** was utilized, containing the $T_c$ of oxides and metals extracted from existing studies (33465 records). The reason behind the selection of the file pertains to the subject of the present study, which is focused on LM alloy superconductors. A significant aspect of the file is the presence of numerous superconductor samples comprising metal elements.

Before model training, we need to preprocess the data to get a better training effect. As illustrated in **Table 1**, the data were preprocessed to address some issues, which then led to the resulting dataset, **mdr_clean.csv**.

**Table 1. Problems and processing methods of SuperCon dataset**

| Problems | Methods |
|---|---|
| (1) Abnormal data | Delete; Correct |
| (2) Data format inconsistency | Uniform format |
| (3) Insufficient data | Replenishment |

## (1) Abnormal data

a. If the $T_c$ corresponding to the formula is missing, the data is deleted and a total of 7107 pieces of data are processed. (marked **Tc_lost** in the dataset)

b. If the formula subscript contains an unknown value (x, y, z), the data is deleted

and a total of 5356 data items are processed. (marked **unknow_num** in the dataset)

c. If the data is misleading for the model, three cases are considered here. (i) The article from which the data originates is withdrawn or there are no corresponding data; (ii) The $T_c$ belongs to a high pressure superconductor; (iii) No data source. If the above situation occurs, the data will be deleted and a total of 54 data will be processed. (Labeled in the dataset as i. **wrong_data**, ii. **high_pressure**, iii. **no_source**)

d. After screening, it is found that the symbols and occurrences outside the periodic table are (Ah: 1, Bf: 1, Ct: 3, M, D, T: 327, Dc: 1, Gs: 1, Jr: 1, Ls: 1, Pf: 1, Ph: 1, Rr: 1). If the formula contains the above element symbols, the data will be deleted and a total of 339 elements will be processed. (marked as **illegal_element** in the dataset).

e. Incorrect data has been entered, the formula or $T_c$ has been changed and a total of 6 data elements have been processed. (Marked **Available_Rev** in the dataset)

A total of 7107 + 5356 + 54 + 339 + 6 = 12862 data were processed and 12856 data were deleted.

**(2) Data format inconsistency**

Taking $Nb_{0.44}Ti_{0.56}$[7] and $Nb_{75}Ga_{7.5}Al_{17.5}$[8] as examples, in the dataset, the subscript numbers used for the alloy formula are atomic quantity ratio (at) and mass percentage (wt%), respectively. As the sum of the subscripts of the alloy formula (expressed by mass percentage) is 100, which is considerably larger than the subscript values of other formulas, it is necessary to avoid the influence of the size of the subscript number. With regard to the characteristics of the dataset, the subscript of the molecular formula $E_{w_1}^1 E_{w_2}^2 \dots E_{w_n}^n$ expressed in terms of mass percentage is converted to an atomic quantity ratio formula $E_{a_1}^1 E_{a_2}^2 \dots E_{a_n}^n$ with a sum of 1 by atomic mass, (e.g. $Nb_{75}Ga_{7.5}Al_{17.5}$ to $Nb_{0.516}Ga_{0.069}Al_{0.415}$) achieved by **Equation (1)**:

$$a_i = \frac{w_i}{W_i} / \sum_i^n \frac{w_i}{W_i} \tag{1}$$

where $E^i$ represents the "$i$-th" element, while $W_i$ denotes the atomic mass of the element $E^i$. The subscript $a_i$ and $w_i$ represent the atomic quantity ratio and mass percentage of the element $E^i$ in relation to the formula. The conversion of the sum of the atomic weight ratios to 1 allows for the consideration of alloys as analogous to monomers.

According to the above method, a total of 350 pieces of data are processed.

**(3) Insufficient data**

As summarized by Konno[9] and Hosono et al.[10], 386 non-superconductors with $T_c$ =0 were added to the SuperCon dataset in this study.

After processing the above three problems, the obtained data features are shown in **Figure 2**, and 20,995 data points remain. **Figure 2(a)** shows the distribution of different $T_c$ values in the dataset. It can be seen that the distribution is uneven and the superconductor data below 40 K is the main data, indicating that the model may be better at predicting the superconductor below 40 K on this dataset. At present, the mainstream superconductoors are usually binary compounds to five-membered compounds, which can be easily seen from **Figure 2(b)**. Considering that some superconductors may not have superconducting properties, we choose the data with $T_c$ above 5 K. In **Figure 2(c)**, the number of occurrences of different elements in these data is counted to illustrate the research popularity of different elements and their potential in superconducting materials.

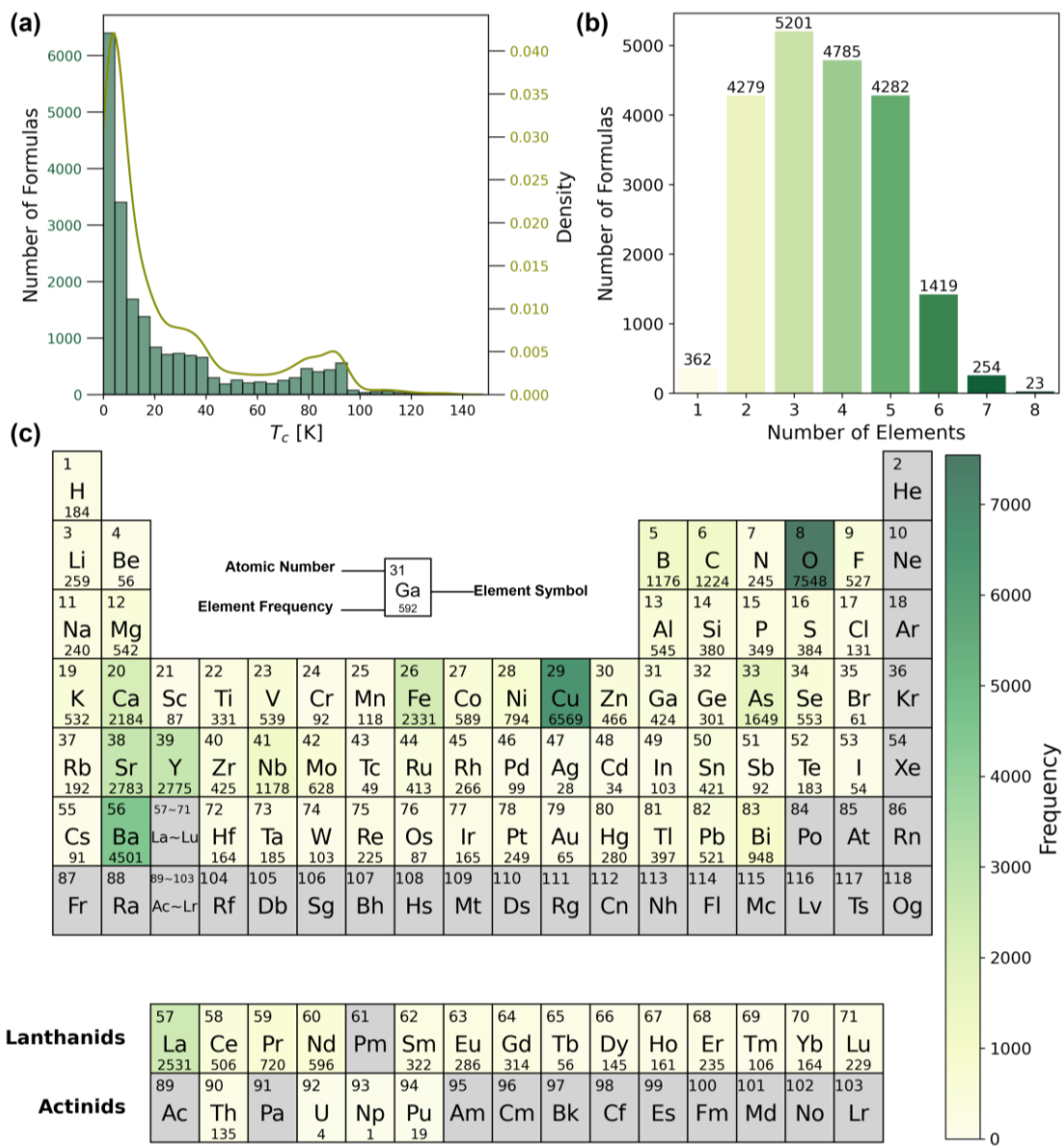

**Figure 2. Features of the mdr_clean.csv dataset: (a) Quantity distribution of $T_c$ values; (b) Number distribution of formula with different numbers of elements; (c) Frequency of occurrence of different elements in the dataset ($T_c > 5$ K).**

## 2) Model comparison

After processing the dataset, we begin to select the appropriate model to train. Several machine learning models are used to predict $T_c$ on the SuperCon dataset. **Table 2** shows the models and feature descriptors used in recent studies, and the effectiveness of the models is evaluated from three aspects: $R^2$, RMSE and MAE. The number of features used is provided in parentheses.

**Table 2. A summary of machine learning model for predicting $T_c$ based on SuperCon dataset.**

| Type | Algorithm | Feature | Feature database | $R^2$ | RMSE(K) | MAE(K) |
|---|---|---|---|---|---|---|
| Tree Based Model | Random forest[11] | Elemental properties (145) | Magpie | 0.885 | - | - |
| | XGBoost[12] | Elemental properties (81) | ElementData in Mathematica | 0.92 | 9.5 | - |
| | Random forest[13] | Elemental properties (53) | Materials Project, Pymatgen, Mendeleev | 0.92 | - | - |
| | Bagged tree[14] | Composition (-) | One-hot formula | 0.93 | 8.91 | - |
| | XGBoost[15] | ec (18) | Magpie | 0.877 | - | 5.4 |
| | | ec+ (23) | Magpie | 0.913 | - | 4.5 |
| | | ec++ (28) | Magpie | 0.921 | - | 4.2 |
| | CatBoost[16] | Elemental properties (322) | Jabir | 0.952 | 6.5 | - |
| | **ExtraTrees (This work)** | **Composition (-)** | **One-hot formula** | **0.952** | **6.26** | - |
| Neural Network Model | CNN+LSTM[17] | Atom vector (20) | SVD in one-hot formula | 0.899 | 83.565 | 5.023 |
| | | Composition (-) | one-hot formula | 0.899 | 83.565 | 5.023 |
| | | Elemental properties (-) | Magpie | 0.899 | 83.565 | 5.023 |
| | CNN[9] | Periodic table (4*32*7) | Periodic table, one-hot formula | 0.92 | - | - |
| | CGCNN[18] | Crystal graph (-) | Materials Project | 0.92 | - | 5.6 |
| | CNN[19] | Elemental properties (81) & Composition (86) & lattice parameter(8) | Pymatgen, Materials Project, one-hot formula | 0.9429 | - | - |
| | DNN[20] | Composition (-) | One-hot formula | 0.95 | 6.08 | 3.08 |
| | | Elemental properties (-) | Magpie, mendeleev, villars | 0.93 | 7.35 | 3.73 |
| | BNAS[21] | Electron band structure (18*32*32*32) | DFT calculation | 0.918 | - | - |
| Integrated Model | CNN+GBDT[22] | Elemental properties (22*6) | Matminer | 0.937 | 4.653 | 8.695 |
| | Deep forest[23] | Composition (86) | One-hot formula | 0.945 | 5.51 | 4.04 |
| | CNN+XGBoost[24] | Elemental properties (81) & Composition (86) | ElementData in Mathematica, one-hot formula | 0.934 | 8.7915 | - |

Tabular data is defined as data that is organized in a tabular form, where the data is arranged into rows and columns, with each row representing a data record and each column representing a feature or attribute. This data structure is ubiquitous in statistics and databases, and is typically employed to store structured data. The data utilized in the study, **mdr_clean.csv**, falls within this category.

Grinsztajn et al.[25] posit that tree-based models (e.g., XGBoost and Random Forest) display certain advantages when dealing with tabular data. (1) **Maintain the orientation of the data**: Since tree-based models are not rotationally invariant, they do not alter the orientation of the features during training and testing. This property enables the model to capture the inherent structure of the data and the actual relationship between features, as opposed to relying on a linear combination of features. (2) **Learn irregular functions**: Objective functions in tabular data may contain irregular patterns, and tree-based models demonstrate a superior ability to learn these patterns. In contrast, neural networks tend to exhibit a preference for smooth solutions and encounter greater difficulty in learning irregular patterns in the objective function. (3) **Less computational cost**: Tree-based models typically exhibit superior training speed compared to deep learning models, particularly when the data set is not overly extensive. (4) **Do not need complex regularization**: Proper regularization and careful optimization can enable neural networks to learn irregular patterns. However, tree-based models can perform well even without complex regularization. They reveal that tree-based models continue to represent a state-of-the-art in medium-sized datasets (~10,000 samples), even in the absence of consideration for their superior training speed. These advantages substantiate the preeminence of tree models as the optimal tool for the processing of flat data in numerous practical applications.

As can be seen from the comparison in **Table 2**, the performance of the tree-based model is almost equal to that of the popular neural network model. Combined with Grinsztajn's theory, our study decides to use the tree-based model as a prediction tool. As demonstrated in 错误!书签自引用无效。, the **mdr_clean.csv** dataset is divided as the training set and the test set at a ratio of 9:1, and eight tree-based models are trained with the default training parameters. Following a comparison of RMSE, $R^2$ and the accuracy of the test set, it was determined that Extra Trees (ET) model and Random Forest (RF) model had a superior effect. Consequently, the hyperparameter optimization of these two models was employed as the prediction model.

To be noted that, we developed the tree-based models in this study using scikit-learn[26], CatBoost[27], XGBoost[28], lightGBM[29], which are powerful and efficient machine learning Python libraries. We also used the grid search function in scikit-learn to generate candidate models from a grid of parameter values and compare them to obtain the optimal model, the work can be seen in Section: **4) Hyperparameter optimization**.

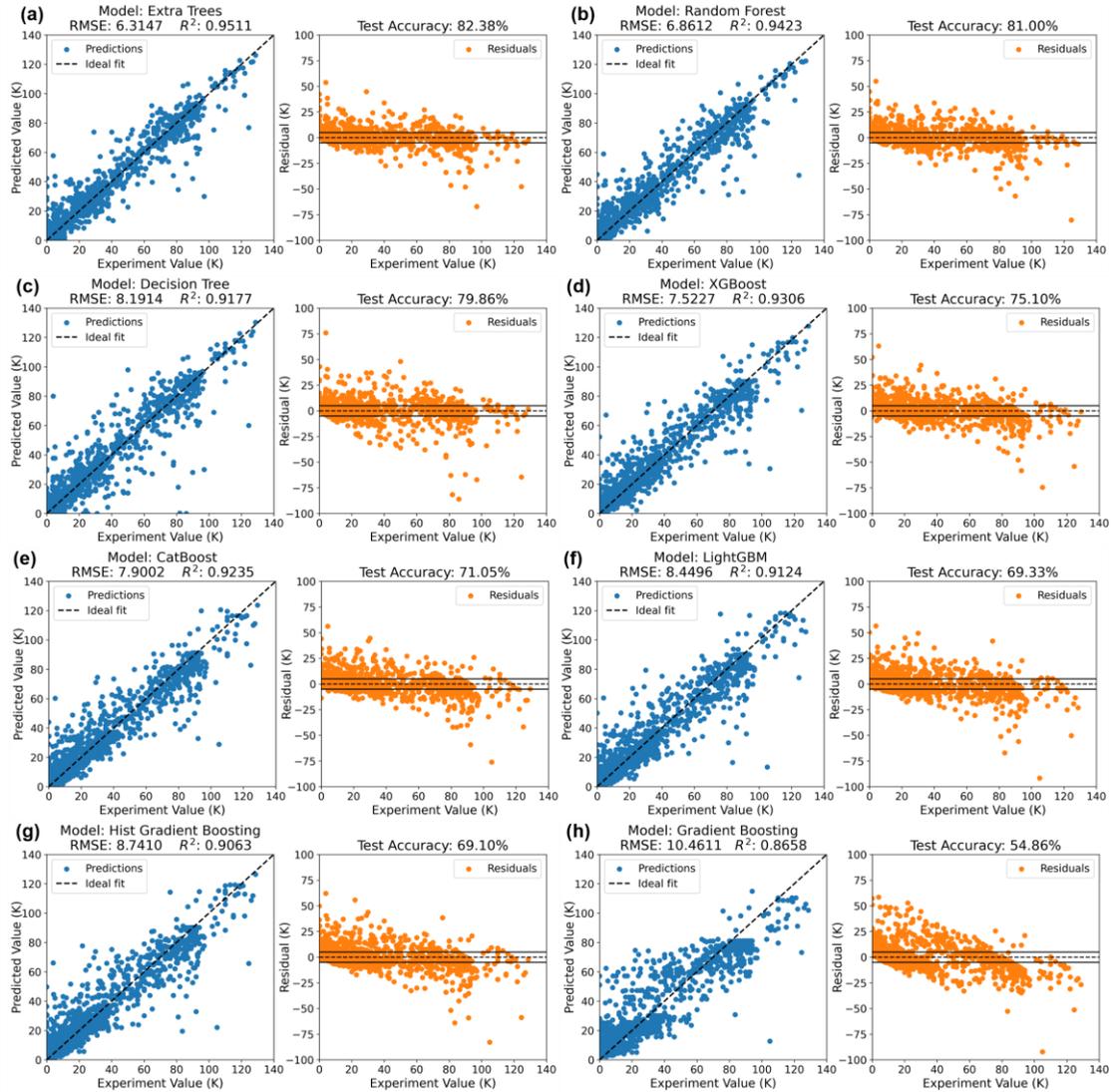

**Figure 3. Performance of tree-based regression models in predict T$_c$. Left: Comparison of predicted and experimental values on the validation dataset. Right: The difference between the predicted value and the experimental value on the validation set, 5K is taken as the prediction threshold and the prediction is considered accurate if the difference is less than 5K.**

### 3) Model evaluation

To better understand the performance of a model, three indicators used in this study are introduced: RMSE, R$^2$, and accuracy.

### (1) RMSE

In tree-based regression models, two distinct loss functions are employed for the purpose of quantifying the discrepancy between the model's predicted value and the actual value. These are the mean squared error (MSE) and Friedman MSE.

The hyperparameter 'squared error' in code is used to refer to the concept of MSE, which is achieved by calculating the square of the difference between the predicted value and the actual value. This can be expressed as follows:

$$MSE = \frac{1}{n}\sum_{i=1}^{n}(y_i - \hat{y}_i)^2 \tag{2}$$

where $y_i$ is the actual value, $\hat{y}_i$ is the predicted value, and $n$ is the total sample. Because MSE is a squared error operation, MSE has a greater penalty for large errors and is insensitive to small errors.

The hyperparameter 'friedman_mse' is an improvement of MSE based on Friedman's decision tree model, which not only considers the square of error but also the probability of event occurrence to revise the classification standard. This method measures the performance of the model by calculating the average error of the left and right subtrees and the number of samples. The calculation equation is as follows:

$$Friedman\_MSE = \frac{n_l n_r (mean_l - mean_r)^2}{n_l + n_r} \tag{3}$$

where $n\_l$ and $n\_r$ are the number of samples of the left and right subtrees respectively, and $mean\_l$ and $mean\_r$ are the average errors of the left and right subtrees respectively. Compared to MSE, Friedman_MSE considers the inhomogeneity of the sample distribution and is more sensitive to small error variations.

RMSE is the square root of the MSE:

$$RMSE = \sqrt{MSE} \tag{4}$$

**(2) $R^2$**

$R^2$, also known as the coefficient of determination, is a statistic in regression analysis that measures how well a model fits the observed data. Its value is between 0 and 1, and the closer it is to 1, the better the model interprets the data. This can be expressed as follows:

$$R^2 = 1 - \frac{\sum_{i=1}^{n}(y_i - \bar{y})^2}{\sum_{i=1}^{n}(y_i - \hat{y}_i)^2} \tag{5}$$

where $y_i$ is the actual value, $\bar{y}$ is the mean value, $\hat{y}_i$ is the predicted value, and $n$ is the total sample.

**(3) Accuracy**

In order to measure the accuracy of the prediction, the data whose absolute error (MAE) of prediction is less than 5K is considered to be correct. The accuracy can be obtained as:

$$MAE = |y_i - \hat{y}_i| \tag{6}$$

$$Accuracy = \frac{prediction_{MAE<5}}{prediction_{all}} \times 100\% \tag{7}$$

where $y_i$ is the actual value, $\hat{y}_i$ is the predicted value, $prediction_{MAE<5}$ is the number of the data whose MAE of prediction is less than 5K, and $prediction_{all}$ is all the number of the data in prediction.

**4) Hyperparameter optimization**

Before training through the tree-based model, hyperparameter optimization is required because it can significantly improve the performance and generalization ability of the model: (1) Hyperparameter optimization helps to prevent overfitting and underfitting, and can effectively control the complexity of the model by adjusting parameters. (2) Hyperparameter optimization can improve the prediction accuracy and stability of the model, making it perform better in different dataset. (3) Hyperparameter optimization can also reduce training time and computational resources, for example, by limiting the depth of the tree or reducing the number of trees to reduce computational complexity. (4) Appropriate hyperparameter settings can make the model more concise and easy to explain, which is particularly important in application scenarios where the model needs to be explained and understood.

In conclusion, hyperparameter optimization is a key step in building efficient and reliable tree models, which can significantly improve the overall performance and applicability of the models. We construct the hyperparameter search scope shown in **Table 3**.

**Table 3. Hyperparameter optimization range for tree-based models**

| Hyperparameter | Optimization range |
|---|---|
| n_estimators | [100, 200, 300, 400, 500] |
| max_depth | [none, 10, 20] |
| min_samples_split | [2, 5, 10] |
| min_samples_leaf | [1, 2, 4] |
| criterion | ['squared_error', 'friedman_mse'] |

Following the conclusion of the hyperparameter search, the MSE values are subjected to a five-fold cross-validation of hyperparameters, with the values listed in descending order. Tables 4 and 5 present a selection of results from the hyperparameter search, with the optimal hyperparameters for the two models highlighted. Next, we will use the two sets of hyperparameters highlighted to train the two models separately.

**Table 4. The result of hyperparameter optimization for RF models**

| hp_a | hp_b | hp_c | hp_d | hp_e | mean_RMSE (K) | RMSE (K) | $R^2$ | Accuracy | Time (s) |
|---|---|---|---|---|---|---|---|---|---|
| F | None | 1 | 2 | 300 | 7.6488 | 7.5649 | 0.9289 | 80.11% | 152.07 |
| S | None | 1 | 2 | 500 | 7.6492 | 7.5747 | 0.9287 | 80.19% | 155.31 |
| S | None | 1 | 2 | 400 | 7.6503 | 7.5658 | 0.9289 | 80.21% | 233.90 |
| S | None | 1 | 2 | 300 | 7.6503 | 7.5630 | 0.9290 | 80.09% | 228.37 |
| F | None | 1 | 2 | 500 | 7.6513 | 7.5849 | 0.9286 | 80.21% | 76.17 |
| F | None | 1 | 2 | 400 | 7.6522 | 7.5764 | 0.9287 | 80.11% | 309.43 |
| F | None | 1 | 2 | 200 | 7.6535 | 7.5767 | 0.9287 | 80.26% | 78.10 |
| S | None | 1 | 2 | 200 | 7.6565 | 7.5723 | 0.9288 | 80.30% | 385.69 |
| F | None | 1 | 2 | 100 | 7.6739 | 7.5385 | 0.9294 | 80.04% | 305.88 |
| S | None | 1 | 5 | 500 | 7.6803 | 7.6613 | 0.9271 | 79.69% | 383.52 |
| … | … | … | … | … | … | … | … | … | … |

**Table 5. The result of hyperparameter optimization for ET models**

| hp_a | hp_b | hp_c | hp_d | hp_e | mean_RMSE (K) | RMSE (K) | $R^2$ | Accuracy | Time (s) |
|---|---|---|---|---|---|---|---|---|---|
| F | None | 1 | 5 | 300 | 7.1686 | 6.2624 | 0.9519 | 81.90% | 227.40 |
| F | None | 1 | 5 | 500 | 7.1721 | 6.2446 | 0.9522 | 81.86% | 379.10 |
| S | None | 1 | 5 | 500 | 7.1743 | 6.2381 | 0.9523 | 81.71% | 380.61 |
| F | None | 1 | 5 | 400 | 7.1745 | 6.2345 | 0.9523 | 81.86% | 303.44 |
| S | None | 1 | 5 | 400 | 7.1762 | 6.2522 | 0.9521 | 81.90% | 305.11 |
| S | None | 1 | 5 | 300 | 7.1789 | 6.2526 | 0.9521 | 81.76% | 228.61 |
| F | None | 1 | 5 | 200 | 7.1795 | 6.2489 | 0.9521 | 81.95% | 151.47 |
| S | None | 1 | 5 | 200 | 7.1797 | 6.2621 | 0.9519 | 81.67% | 152.31 |
| S | None | 1 | 5 | 100 | 7.1890 | 6.3009 | 0.9513 | 81.57% | 76.10 |
| F | None | 1 | 5 | 100 | 7.1958 | 6.2923 | 0.9514 | 81.62% | 75.59 |
| … | … | … | … | … | … | … | … | … | … |

Note: hp_a, hp_b, hp_c, hp_d, hp_e correspond to five hyperparameter categories, which are respectively given below:

hp_a: criterion, hp_b: max_depth, hp_c: min_samples_leaf, hp_d: min_samples_split, hp_e: n_estimators.

F and S mean two methods of calculating the loss function:

F: Friedman_mse, S: squared_error.

## 3. Results

The strategy employed in this study to construct the formula of alloys involves the following steps (shown in **Figure 4(a)**): The given elements are designated as an input layer, and the atomic proportion is assigned to each element as a weight. This is then fed into an intermediate layer, each of which has a different processing logic to select the number of elements fed in and to ensure that the sum of the weights assigned as atomic proportions is one. The formula of alloys with the corresponding number of elements are then output. The output formula is then fed into the prediction model and the $T_c$ is obtained.

In this study, eight metal elements [Ga, Bi, In, Sn, Zn, Ag, Sb, Cu] are extracted according to the LM alloys summarized by Zhang et al.[30] for printed electronics, and $T_c$ of the binary and ternary alloys composed of these elements is predicted. To ascertain the proportion with the highest $T_c$ in a family of alloys at the lowest computational cost, **Figure 4(b)** demonstrates the influence of the number of different weights on the prediction. It is observed that when the number exceeds 1000, the maximum $T_c$ is stable at 7.01 K, the corresponding alloy is also stable at $In_{0.5}Sn_{0.5}$, and the second and third $T_c$ values are also stable. Therefore, it is concluded that 1k random samples can be used to search for the highest $T_c$ of binary and ternary alloys.

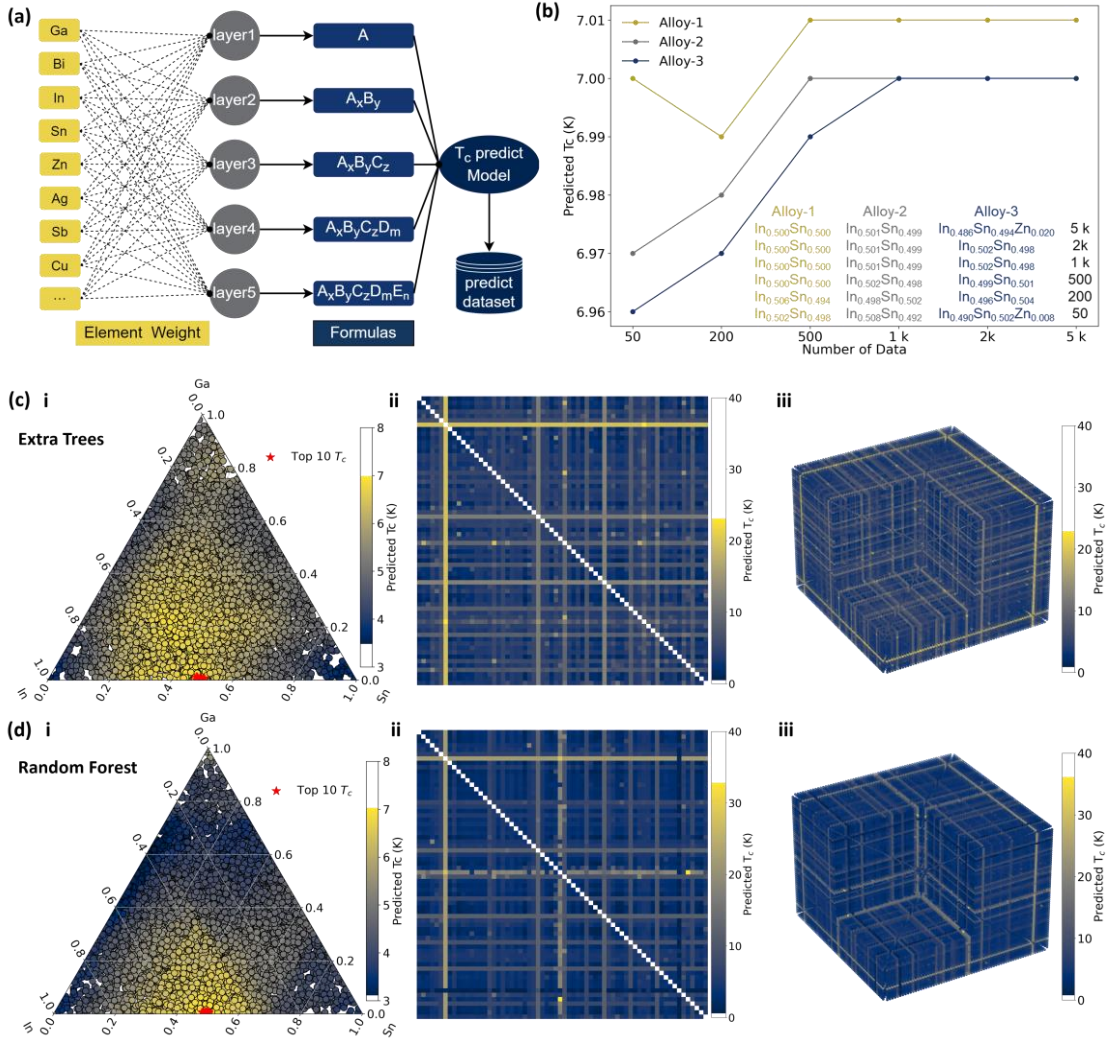

**Figure 4. Finding LM-Alloy with optimum T_c. (a) T_c prediction process of LM-Alloy; (b) Impact of the amount of data on the prediction effect of ternary alloys. (c) Extra Trees Model and (d) Random Forest Model: (i) Predicted T_c distribution of Ga-In-Sn alloys with 10000 different components. (ii) Predicted T_c for 2145 binary alloys. (iii) Predicted T_c for 45760 ternary alloys.**

As demonstrated in **Figure 4(c.i)** and **Figure 4(d.i)**, 10,000 random samples are obtained for the Ga-In-Sn alloy, and two models (Extra Trees and Random Forest) are utilized for the prediction of $T_c$. The $T_c$ distribution corresponding to the various samples is thus obtained, with the ten sample points with the highest $T_c$ marked with red stars, as illustrated in **Figure 4(b)**. This outcome underscores the efficacy and cost-effectiveness of prediction with a limited sample size of 1k in binary and ternary alloys.

LM generally refers to the metals that have a melting point at or near room temperature, but in a broad sense, metals can become liquid within a certain temperature zone. This study attempts to find more LM alloy superconductors outside the conventional temperature zone. 66 kinds of metal elements [Ag, Al, Au, Ba, Be, Bi, Ca, Cd, Ce, Co, Cr, Cs, Cu, Dy, Er, Eu, Fe, Ga, Gd, Ge, Hf, Hg, Ho, In, Ir, K, La, Li, Lu,

Mg, Mn, Mo, Na, Nb, Nd, Ni, Np, Os, Pb, Pd, Pr, Pt, Pu, Rb, Re, Rh, Ru, Sb, Sc, Sm, Sn, Sr, Ta, Tb, Tc, Th, Ti, Tl, Tm, U, V, W, Y, Yb, Zn, Zr] were extracted from **mdr_clean.csv** and the combination of binary total and ternary alloys are carried out. According to the conclusions above, 1k samples are taken for each alloy family. The $T_c$ is predicted by two models (Extra Trees and Random Forest) and the highest $T_c$ is recorded. **Figure 4(c.ii)** and **Figure 4(d.ii)** are obtained by taking 66 metals as the horizontal and vertical axes (the order of the elements in the horizontal axis is from left to right, and the order of the elements in the vertical axis is from top to bottom, and the Figure with element coordinate axes can be further shown in more supplementary figures, and the highest $T_c$ of the corresponding binary alloy can be mapped as the color. It should be noted that the binary alloys involved in the figure do not necessarily exist, and the corresponding predicted values are for reference only. In the same way, the highest $T_c$ of 45,760 ternary alloys was also recorded and plotted in **Figure 4(c.iii)** and **Figure 4(d.iii)**.

**Figure 5(a,b)** shows the performance of both models on the training set. It is imperative to acknowledge the pivotal role of the formula in determining the $T_c$. It is recognized that $T_c$ in **mdr_clean.csv** dataset varies considerably between study (seen from **Figure 5(c)**), attributable to various factors, including but not limited to differences in atomic structure, material size, and environmental conditions. Consequently, the training set is unable to achieve a perfect regression.

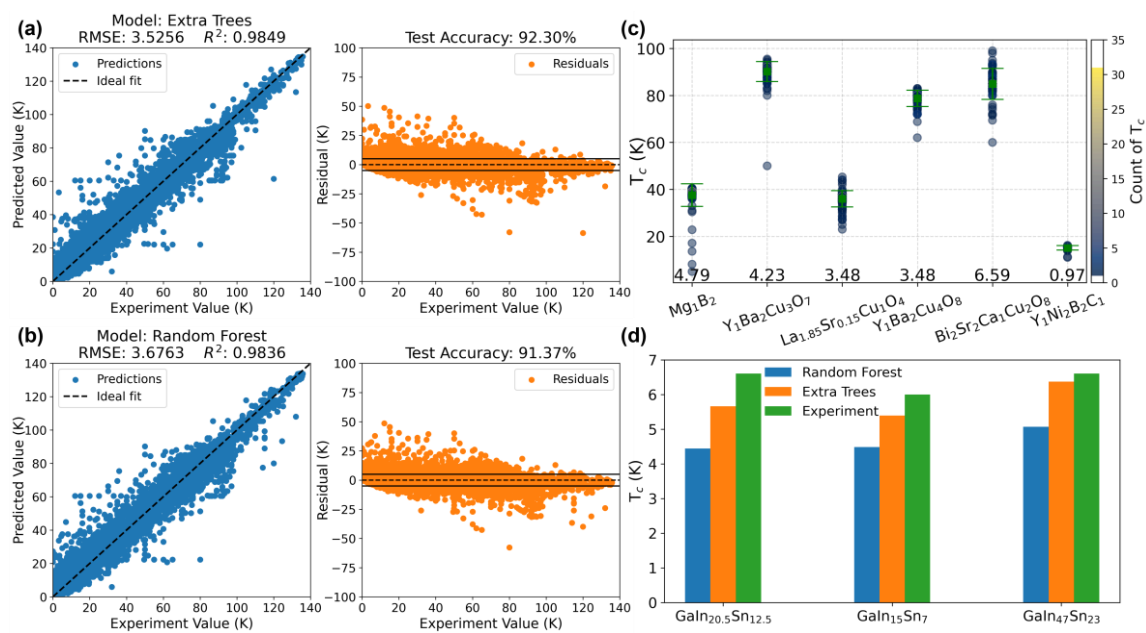

**Figure 5. Test result of model on training set: (a) Extra Trees Model; (b) Random Forest Model. (c) The distribution and standard deviation of some data in the mdr_clean.csv dataset. The bottom of the graph shows the value of the standard deviation. (d) The $T_c$ of LM-Alloy (GaIn$_{20.5}$Sn$_{12.5}$[1], GaIn$_{15}$Sn$_7$[32], GaIn$_{47}$Sn$_{23}$[31]) is predicted and compared with experimental values in the mdr_clean.csv dataset.**

According to the study of Ren et al.[31], by adjusting the composition of Ga-in-Sn alloys, the maximum $T_c$ can reach 6.6 K, which is close to the predicted results shown in **Figure 4(c.i)** and **Figure 4(d.i)**, verifying the accuracy of Extra Trees model. This study also uses the model to predict the $T_c$ of part of the Ga-In-Sn alloy in the present study, and the prediction results are shown in 错误!未找到引用源。**(d)**. All the errors shown by Extra Trees are within 1 K, and the accuracy standard of this study was 100%.

## 4. Discussion

It is evident that there are still some issues in this study that can be improved from the following perspectives:

**1) Data preprocessing**

In formulas, information about the position of elements is very important. Some elements need to be placed together to form groups to eliminate the effect of molecular isomerism on the expression of the formula. The identification of groups is complicated, taking $Ca(OH)_2$ as an example, it can be identified by parentheses, but when there is only one group, such as NaOH, the parentheses are omitted. However, the one-hot coding used in this study only records the proportion of elements in the formula. In the absence of structural information, fully mining the formula information through data preprocessing is conducive to better training of the model.

**2) Data completion**

(a) The dataset exerts a significant influence on the model, which is vulnerable to erroneous predictions due to an absence of cognition regarding non-superconducting data. It is imperative to incorporate non-superconducting data information. (b) Given that this model is employed for the prediction of the superconducting transition temperature at normal pressure, it is essential to collect corresponding pressure data to expand the scope of model training, in view of the recent emergence of high pressure superconductors. (c) In view of the limited information of formula, the collection of structural information of the material also helps to enhance the performance of the model.

**3) Model reliability**

The reliability of the model has been described by RMSE, $R^2$, Accuracy, which does not indicate that the predictions of the model are always reliable. The comparison of $T_c$ for some Ga-In-Sn alloys suggests that the model developed in this study is a good reference for predicting $T_c$ in LM alloys.

**(4) More prediction**

The increase in the complexity of alloy types, owing to the increase in the number of alloy elements, has resulted in the incomplete prediction of the $T_c$ of alloys with more diverse elements, such as quaternary and above. This is due to limitations in the arithmetic power of the search algorithm. In future studies, the search algorithm will be redesigned to circumvent the aggressive search and expedite the prediction of the $T_c$ of multivariate alloys. Concurrently, future research will augment the prediction types to encompass a more extensive array of metal oxides, leveraging the comprehensive dataset of **mdr_clean.csv**.

## 5. Conclusion

In this study, the current issues with the SuperCon dataset were systemized in order to create a more effective dataset, **mdr_celan.csv**. It was then used to train eight tree-based models for the prediction of $T_c$ from formula. Following a thorough evaluation, it is ascertained that the most efficacious Extra Trees regression model ($R^2 = 0.9519$, RMSE = 6.2624) was employed, complemented by the second most effective Random Forest model ($R^2 = 0.9289$, RMSE = 7.5649). The Extra Trees regression model is identified as the most effective model in the extant study, exhibiting an optimal balance of $R^2$ and RMSE metrics. The Extra Trees regression model is employed to predict the $T_c$ of several printable LM-alloys, with an error range of 1k observed in comparison to experimental results. Among the printable LM-alloys, the $In_{0.5}Sn_{0.5}$ is identified as having the highest $T_c$: 7.01 K, a property that could potentially enable the fabrication of superconducting wires.

The present study also attempted to study LM-alloy superconductors in a broad sense by collecting all the 66 metal elements in **mdr_celan.csv**, which comprise 2145 binary alloys and 45,760 ternary alloys. It is acknowledged that not all of these alloys may be extant; However, all of them have been predicted to exhibit a $T_c$ greater than 0 K and less than 40 K, a range that could prove useful to experimentalists. Overall, given the efficiency and extensibility of the current methodology especially tremendous spaces in material options provided from liquid metal genome[33] and combinatorics theory[34], more candidate superconductors within different working temperature scales can possibly be screened out in the near future.

## Disclosure statement

## Data availability

The data used for drawing the graphs in this work comes from model calculations and can be provided upon request from the author(s).